%% file: main.tex
\documentclass[sigconf]{acmart} 
\copyrightyear{2020}
\acmYear{2020}

\pdfoutput=1

\def\BibTeX{{\rm B\kern-.05em{\sc i\kern-.025em b}\kern-.08emT\kern-.1667em\lower.7ex\hbox{E}\kern-.125emX}}

\usepackage{amssymb}
\usepackage{algorithm}
\usepackage{algorithmic}
\usepackage{bibentry}
\usepackage{amsfonts}
\usepackage{subfigure}
\usepackage{multirow}
\usepackage{amsmath}
\usepackage{graphicx}
\usepackage{epsfig}
\usepackage{color}
\usepackage{epstopdf}
\usepackage{rotating}
\usepackage{caption}
\usepackage{float}
\usepackage{multicol}
\usepackage{amssymb}
\usepackage{graphicx}
\usepackage{bbding}
\usepackage{algorithm}
\usepackage{algorithmic}
\usepackage{balance}
\usepackage{color}
\usepackage{amssymb}
 
\begin{document}

\title{Attacking Black-box Recommendations via \\Copying Cross-domain User Profiles}

\author{Wenqi Fan}
\affiliation{%
  \state{City University of Hong Kong}
}
\email{wenqifan03@gmail.com}

\author{Tyler Derr}
\affiliation{%
  \state{Michigan State University}
}
\email{derrtyle@msu.edu}

\author{Xiangyu Zhao}
\affiliation{%
  \state{Michigan State University}
}
\email{zhaoxi35@msu.edu}

\author{Yao Ma}
\affiliation{%
  \state{Michigan State University}
}
\email{mayao4@msu.edu}

\author{Hui Liu}
\affiliation{%
  \state{Michigan State University}
}
\email{liuhui7@msu.edu}

\author{Jianping Wang}
\affiliation{%
  \state{City University of Hong Kong}
}
\email{	jianwang@cityu.edu.hk}

\author{Jiliang Tang}
\affiliation{%
  \state{Michigan State University}
}
\email{tangjili@msu.edu}

\author{Qing Li}
\affiliation{%
  \state{The Hong Kong Polytechnic University}
}
\email{csqli@comp.polyu.edu.hk}

\begin{abstract}
Recently, recommender systems that aim to suggest personalized lists of items for users to interact with online have drawn a lot of attention. In fact, many of these state-of-the-art techniques have been deep learning based. Recent studies have shown that these deep learning models (in particular for recommendation systems) are vulnerable to attacks, such as data poisoning, which generates users to promote a selected set of items. However, more recently, defense strategies have been developed to detect these generated users with fake profiles. Thus, advanced injection attacks of creating more `realistic' user profiles to promote a set of items is still a key challenge in the domain of deep learning based recommender systems. In this work, we present our framework CopyAttack, which is a reinforcement learning based black-box attack method that harnesses real users from a source domain by copying their profiles into the target domain with the goal of promoting a subset of items. CopyAttack is constructed to both efficiently and effectively learn policy gradient networks that first select, and then further refine/craft, user profiles from the source domain to ultimately copy into the target domain. CopyAttack's goal is to maximize the hit ratio of the targeted items in the Top-$k$ recommendation list of the users in the target domain. We have conducted experiments on two real-world datasets and have empirically verified the effectiveness of our proposed framework and furthermore performed a thorough model analysis. 

\end{abstract}

\keywords{ Recommender Systems;   Cross-Domain; Data Poisoning Attacks, Black-box Attacks}
\maketitle

\input{introduction}

\input{relatedwork}
\input{problem}

\input{model}
\input{experiments}

\input{conclusion}

\bibliographystyle{acm-reference-format}
\balance
\bibliography{main.bbl}
\appendix
\input{appendix.tex}
\end{document}

%% file: introduction.tex
\section{Introduction}

\begin{figure}[t]
\centering
{\includegraphics[width=0.99\linewidth]{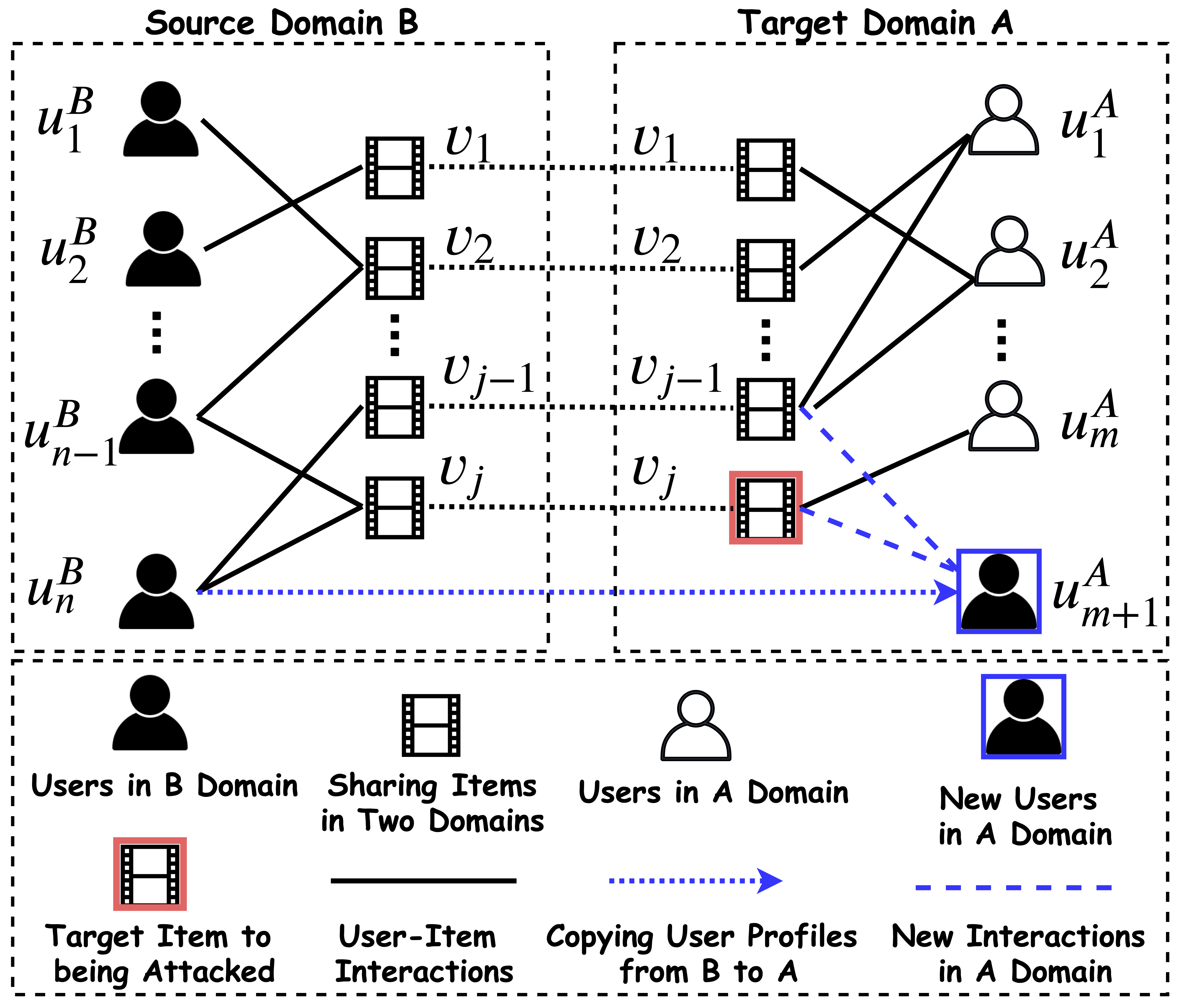}}
\caption{Two domains share some movies. The profile of user $u_n^B$ in the source domain $B$ is copied into the target domain $A$ for attacking the target item $v_j$.} 
\label{fig:overlappingitems}
\end{figure}

Recommender systems aim to suggest a personalized list of items that users are likely to interact with (e.g., click or purchase) in online worlds, especially in many user-oriented online services such as E-commerce (e.g., Amazon and Taobao), and Social Media sites (e.g., Facebook and Twitter). Recent years have witnessed increasing efforts in adopting deep learning techniques such as RNNs and GNNs for recommendations~\cite{wang2019neural}. These deep learning based recommender systems have achieved the state-of-the-art performance. However, it is well known that deep neural networks (DNNs) are highly vulnerable to adversarial attacks~\cite{goodfellow2014explaining,dai2018adversarial,zugner2018adversarial} where adversaries tend to manipulate the data for degrading the prediction performance. Recent studies have demonstrated that the DNNs based recommender systems are also vulnerable to adversarial attacks~\cite{yang2017fake,christakopoulou2019adversarial} where adversaries intend to manipulate users' decisions for their desires. One of the most popular ways to attack recommender systems is data poisoning attacks (also called as shilling attacks)~\cite{li2016data,christakopoulou2019adversarial,chen2018shilling,fang2018poisoning,yang2017fake}. In these attacks, adversaries generate users in a recommender system with well-designed profiles to promote a carefully chosen subset of items~\cite{li2016data, christakopoulou2019adversarial, lam2004shilling}. However, recent defense studies~\cite{zhang2015catch,cai2019detecting,chen2018shilling,wu2012hysad} have demonstrated that these fake profile users are easy to be detected since they present very different patterns from real profiles. Thus, how to inject users with profiles similar to real ones is still a key challenge to attack the DNNs based recommender systems.




Some real-world recommendation platforms have similar functionalities and as a consequence, they have a lot of information in common. For example, movie recommendation platforms IMDb and Netflix share a lot of movies and e-commerce sites Amazon and eBay have millions of products in common. Moreover, users from these platforms with similar functionalities also share similar behavior patterns/preferences. In fact, these observations have encouraged a large body of work targeted on leveraging information from one platform to help recommendations in the other platform that is well known as cross-domain recommendations~\cite{cantador2015cross}. Recall that the key obstacle to attack recommender systems is how to generate users with profiles close to real ones. To tackle this challenge, in this work, we change our perspective -- instead of generating users with fake profiles, we propose to copy cross-domain users with real profiles. One illustrative example is shown in Figure~\ref{fig:overlappingitems}, where we have a target domain $A$ and a source domain $B$ for movie recommendations. These two domains share a set of movies. To attack/promote the targeted item $v_j$ in target domain $A$, user $u_n^B$'s profile $\left \{ v_{j-1}, v_{j} \right \}$ in the source domain $B$ can be copied into the target domain $A$ as a new user $u_{m+1}^A$, such that the movie $v_j$ is attacked.

In this paper, we aim to attack black-box recommendations via copying cross-domain user profiles. The copied user profiles are naturally real. However, how to select user profiles in the source domain under the black-box setting faces tremendous challenges since in the black-box setting, we only have the query access to the target model and each query feedback consists of Top-$k$ recommended items for specific users. Moreover, the majority of existing attack methods have been designed under the white-box setting, in which the attacker requires to have full knowledge of the target model (e.g., model design and parameters) and dataset~\cite{li2016data,welling2011bayesian,christakopoulou2019adversarial}. Existing white-box approaches such as these based on Projected Gradient Method and Stochastic Gradient Langevin Dynamics~\cite{li2016data,christakopoulou2019adversarial} are not applicable to our problem. Therefore, we propose a reinforcement learning (RL) based attack method that learns to choose user profiles in the source domain $B$ with only query feedback from the target recommender systems. Our major contributions are summarized as follows:

\begin{itemize}
\item We introduce a novel strategy to obtain real user profiles by copying cross-domain user profiles to attack the target recommender systems; 

\item We propose a novel framework (\textbf{CopyAttack}) to attack recommendations under the black-box setting via reinforcement learning, which can effectively and efficiently select cross-domain user profiles to perform effective attacks; and 

\item We conduct comprehensive experiments on two real-world datasets to demonstrate the effectiveness of the proposed attacking framework.
\end{itemize}

The remainder of this paper is organized as follows. In Section~\ref{sec:problem} we introduce the problem definition. Thereafter we introduce the proposed framework in Section~\ref{sec:methodlogy}. In Section~\ref{sec:Experiments}, we conduct experiments on two real-world datasets to illustrate the effectiveness of the proposed method. In Section~\ref{sec:relatedwork}, we review related work. Finally, we conclude our work with future directions in Section~\ref{sec:conclusion}.

%% file: relatedwork.tex
\section{Related Work}
\label{sec:relatedwork}

Recommender systems aim to recommend potential items  to specific users.  Attacking recommender systems  can influence users' beliefs  and decisions with malicious purposes~\cite{christakopoulou2019adversarial,lam2004shilling,chen2018shilling}.  Some methods are proposed to study this directions. More specifically, ~\cite{li2016data} apply Projected Gradient Method and Stochastic Gradient Langevin Dynamics (SGLD)~\cite{welling2011bayesian} to optimize data poisoning attack model  with full knowledge of factorization-based collaborative filtering. ~\cite{christakopoulou2019adversarial}  introduces two steps adversarial framework for recommendations, in which they first generate fake users  through Generative Adversarial Networks (GAN), and then apply Projected Gradient Method for further crafting fake user profiles with a suitable adversarial intent. ~\cite{yu2017hybrid} proposed hybrid attacks, which elaborate fake user profiles via fusing ratings information and social relationships for social recommendations. However, many of these data position methods fundamentally rely on the white-box model, in which the attacker requires the adversary to have full knowledge of the target model and dataset~\cite{li2016data}.  That is, they crucially require direct access to the target model, as well as the dataset in recommender systems.  For recommender systems as real-world application scenarios, expecting these kinds of complete access is not realistic. Therefore, it is desired to study black-box attacks in recommender system, where the attackers do not have full knowledge of the target model.   Therefore, we propose a novel framework to attack under black-box setting to fill this gap.

%% file: problem.tex
\section{Problem Statement}
\label{sec:problem}


Let a target recommender system $A$ be defined as having a set of users $\mathcal{U}^A= \left \{u_1^A, u_2^A, ..., u_{n^A}^A \right \}$ and a set of items $ \mathcal{V}^A = \left \{ v_1, v_2, ..., v_{m^A} \right \} $, where $n^A$ is the number of users and $m^A$ is the number of items in $A$.
In addition, user-item interactions are represented as the matrix $\mathbf{Y}^A \in \mathbb{R} ^{n^A \times m^A} $, where an interaction $y^A_{ij}$ indicates that user $u^A_i$ interacted with item $v_j$ (e.g., clicked/bought), and 0 otherwise. Furthermore, we define the set of items a user $u_i^A$ interacts with in $\mathbf{Y}^A$ (i.e., their user profile) as:
\vspace{-0.5ex}
\begin{align}
    P_{u_i}^A = \left \{ v_1 \rightarrow  ... \rightarrow  v_j \rightarrow  ... \rightarrow  v_l  \right \} \nonumber
\end{align}
\vspace{-0.5ex}
where $\rightarrow$ denotes the sequential order of the $l$ items $u_i^A$ has interacted with (and the length $l$ can vary between users). We then denote the set of all user profiles in the target domain $A$ as $\mathcal{P}_{\mathcal{U}}^A =  \left \{  P_{u_1}^A, ..., P^A_{u_i}, ..., P^A_{{u_{n^A}}} \right \}$.

We define the source recommender system $B$ similarly, having the set of users $\mathcal{U}^B$, set of items $\mathcal{V}^B$, interaction matrix $\mathbf{Y}^B \in \mathbb{R}^{n^B \times m^B}$, and set of user profiles $\mathcal{P}_{\mathcal{U}}^B$. Note that the source domain $B$ is selected such that there are overlapping items between the target domain $A$ and source domain $B$. In other words, there exists a set of items $\mathcal{V} = \mathcal{V}^A \cap \mathcal{V}^B$, where $|\mathcal{V}| \neq \emptyset$ and the overlap (i.e., size of $\mathcal{V}$) is assumed to be sufficiently large. Thus, we then define an item profile $P^A_{v_j}$ for $v_j \in \mathcal{V}$, which is the set of users from $A$ who have interacted (e.g., purchased/clicked) with $v_j$ in $\mathbf{Y}^A$ as follows:
\begin{align}
  P_{v_j}^A = \left \{ u^A_1\rightarrow   ...\rightarrow   u^A_i\rightarrow   ...\rightarrow   u^A_o  \right \} \nonumber
\end{align}
where $o$ is the number of user's in the items profile (that can differ from item to item). Let $\mathcal{P}_{\mathcal{V}}^A = \left \{   P_{v_1}, ..., P_{v_j}, ...,  P_{v_{m^A}} \right \} $ denote the set of item profiles in target domain $A$.

Now, given the notations of the target and source recommender systems $A$ and $B$, respectively, we formally define the goal of the target recommender system $A$. Overall, the objective of $A$ (which we denote here at $Rec(\cdot,\cdot)$) is to predict whether user $u_i^A$ likes (i.e., will interact with) an item $v_j$ as $y^A_{ij} = Rec(P^A_{u_i}, P_{v_j}^A )$. Thus, without loss of generality, the target recommender system task is to predict a list of Top-$k$ ranked potential items for each user. More formally, this recommendation is as follows:
\begin{align}
    y^A_{i,>k} = \left \{ v_{[1]} , v_{[2]}, ..., v_{[k]} \right \}  = Rec (P^A_{u_i}, \mathcal{P}_{\mathcal{V}}^A) \nonumber
\end{align}
where $y^A_{i,>k} = \left \{ v_{[1]} , v_{[2]}, ..., v_{[k]} \right \}$ denotes the Top-$k$ candidate items for user $u^A_i$. For completeness, we note that these  candidate items in $y^A_{i,>k}$ are ranked by $Rec(\cdot,\cdot)$, where user $u^A_i$ is more likely to click/purchase item $v_{[i]}$ than $v_{[i+1]}$.

Finally, we define the problem of a black-box injection attack to promote a target item $v_* \in \mathcal{V}$ by copying a set of users $\mathcal{U}^{B \rightarrow A} = \{u_i^B \}_{i=1}^{\triangle}$ from the source domain to the target domain, where $\triangle$ is the budget given to the attacker (in terms of the number of profiles to copy). Note that these results in the target domain having the set of polluted users $\mathcal{U}^{{A}'} = \mathcal{U}^{A} \cup \mathcal{U}^{B \rightarrow A}$ and thus also polluting the interaction matrix $\mathbf{Y}^{A}$. More precisely, the pollution of $\mathbf{Y}^{A}$ is due to the fact that introducing the copied users brings their interactions with the set of items $\mathcal{V}$ and hence disrupts the relations between users and items in $A$. Furthermore, to be more specific, we define the promotion of a target item $v_*$ as having this item appear in the Top-$k$ recommendation list for users in $\mathcal{U}^A$ that previously (before injecting the copied users $\mathcal{U}^{B \rightarrow A}$ and their associated interactions) did not have $v_*$ in their Top-$k$ recommendation list.

%% file: model.tex
\section{The Proposed Framework}
\label{sec:methodlogy}
In this section, we will first give an overview of the proposed framework, then provide details for each of the frameworks components, and finally discuss how to learn the model parameters.

\begin{figure*}[t]
\centering
{\includegraphics[width=0.99\linewidth]{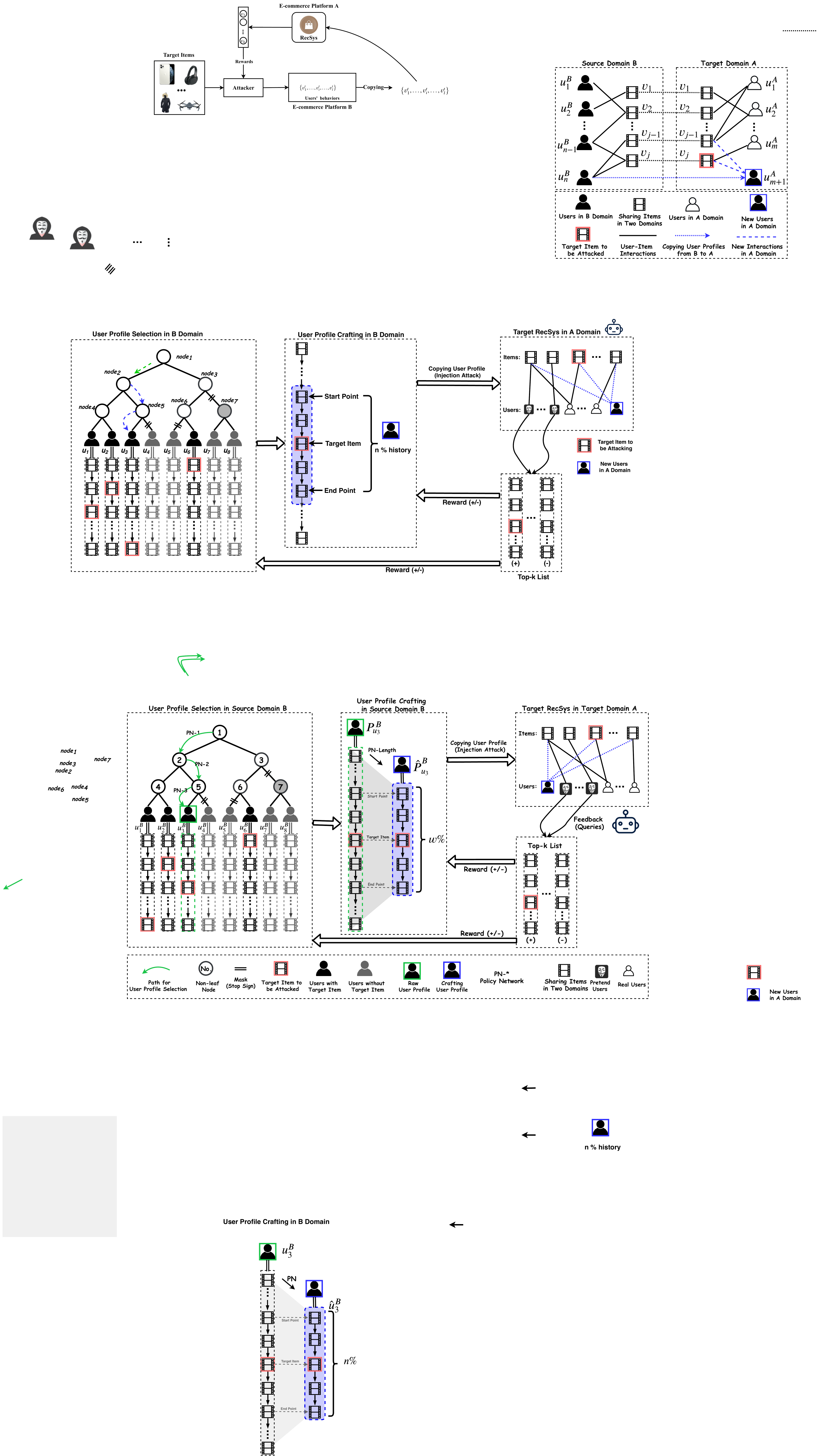}}
\caption{An overview of the proposed framework. It contains three major components: user profile selection in B domain, user profile crafting, and Injection Attack and queries.  }\label{fig:framework}
\end{figure*}

\subsection{An Overview of the Proposed Framework}

To perform attacking in recommender systems in the black-box setting, traditional gradient-based methods~\cite{li2016data,christakopoulou2019adversarial} are not applicable. Thus, we propose a reinforcement learning (RL) based attack method, CopyAttack, to learn the strategy of copying cross-domain user profiles. This is because reinforcement learning provides a natural way to interact with a black-box recommender system. The architecture of CopyAttack is shown in Figure~\ref{fig:framework}, which consists of three major components: user profile selection, user profile crafting, and injection attack and queries. 

The first component is to perform user profile selection for specific target item attack, which is proposed to select user profiles from $\mathcal{P}^{B}_{\mathcal{U}}$ (i.e., user profiles from the source domain $B$). This can be seen in the left part of Figure~\ref{fig:framework}. However, modeling this process of selection with reinforcement learning technique is rather challenging  under limited resources (i.e., number of queries (or interactions) allowed to the target recommender system), since a huge number of user profiles (discrete action space) in source domain $B$ might lead to inefficiency and ineffectiveness at the same time.  Moreover, not all the user profiles are useful to help attack the specific target item in the target recommender system. 
To address these challenges, we propose to adopt hierarchical-structure policy gradient~\cite{arulkumaran2017deep,chen2019large,sutton1998introduction,williams1992simple} with masking mechanism to efficiently learn the strategy of effectively selecting cross-domain user profiles, so as to maximize long-run rewards. 


Next, once having selected a user profile from the first component, the second component is used for profile crafting. Here profile crafting aims to further modify the user profile by considering the reduction of attack cost and can be seen in the center part of Figure~\ref{fig:framework}. We note that users can have user profiles consisting of varying lengths (i.e., number of items they have interacted with). Thus, it could be the case that not all the interactions that the user has given towards items in their user profile are helpful. Furthermore, too long of a user profile might include some noise as well as increase the attack cost (i.e., number of interactions the copied user would need to perform in the target domain). Hence, we introduce a second step policy gradient network to craft the the user profiles by considering this attacking cost issue. More specifically, this second step policy gradient network will decide what percentage of the user profile is kept around the target item $v_*$. 

Lastly, the third component's first objective is to attack the target recommender system by copying the crafted cross-domain user profiles (i.e., those coming from the source domain). After having copied the crafted cross-domain user profile, queries on the target recommender system are performed to obtain some feedback in the form of Top-$k$ recommendations. This feedback is then used to form a reward for optimizing the whole framework (i.e., updating the policy gradient networks of the first and second components). This component can be seen in the right part of Figure~\ref{fig:framework}. 

Next, we will discuss an overview of the attacking environment of our black-box reinforcement learning based attacking method. 

\subsection{Attacking Environment Overview}
The attacking black-box framework can be modeled as Markov Decision Process (MDP)~\cite{gosavi2009reinforcement}. The definition of the MDP contains the state space $S$, action set $A$, transition probability $P$, reward $R$, and discount factor $\gamma$ (i.e., $(S, A, P, R, \gamma)$ ) that are defined as follows:

\textbf{State $S$.} A state $s_t$  consists of all the intermediate injected user profiles at $t$.


\textbf{Action $A$.} The action has two components and is defined as $A= \left \{ a_t =(a^u_t, a^l_t) \right \}$. More specifically, the attacker is allowed to first select a user $a^u_t=u^B_{i}$ from the cross-domain (i.e., source domain) system $B$ at state $t$. Then, the attacker can modify the original profile $P_{u_i}^B$ of $u_i^B$ to craft a profile of perhaps shorter length resulting in $a^l_t = \hat{P}_{u_i}^B$. Note that this crafted user profile would be the one ultimately injected into the target recommender system.


\textbf{Transition probability $P$.} Transition probability $p(s_{t+1} |s_t, a_t)$ defines the probability of state transition from the current $s_t$ to the next state $s_{t+1}$ when the attacker takes action $a_t$. 


\textbf{Reward $R$.} The goal of the attacker is to attack a target item $v_*$ in the target recommender system $Rec(\cdot, \cdot)$ with their desires (such as promotion/demotion of that target item). In this work, we focus on the promotion attack, where the attacker seeks to have the target items recommended to as many users as possible. A natural way to define the reward for the RL based method is on the basis of ranking evaluation measures. We note that this type of reward function based on ranking evaluation is quite general and could be used for either a promotion or demotion attack. Thus, for the reward function based on ranking, we assign a positive reward for action $a_t$ when the target item $v_*$ belongs to the Top-$K$ recommended list for users $u^A_{i*} \in \mathcal{U}_*^A \subset \mathcal{U}^A$. More specifically, the set of users $\mathcal{U}_*^A$ is a set of pretend users that the attacker had already established in the target domain before the injection attacks (as seen in Figure~\ref{fig:framework}). We note that these pretend users solely exist in the target recommender system so that the attacker can use them as a proxy for determining how effective their copied user profiles are are at promoting the target items to all users in $\mathcal{U}^A$. We use the Hit Ratio (HR@K) as the ranking evaluation in our reward function $r(s_t, a_t)$ for a given state $s_t$ and action $a_t$, which we define as follows: 
\vspace{-0.75ex}
\begin{align}\label{eq:reward}
    r(s_t, a_t) = \frac{1}{|\mathcal{U}_*^A|} \sum\limits_{i=1}^{|\mathcal{U}_*^A|} HR(u^A_{i*}, v_*, k) \\
HR(u^A_{i*}, v_*, k) = \left\{\begin{matrix}
1, & v_* \in y_{u^*,>k}, \\ 
0, & v_*  \notin  y_{u^*,>k} \nonumber
\end{matrix}\right.
\end{align}
\vspace{-0.75ex}
where $HR(u^A_{i*}, v_*, k)$ returns the hit ratio for a targeted item $v_*$ in the Top-$k$ listing of the attackers pretend user $u^A_{i*}$ (i.e., whether  $v_*$ is in the set $y^A_{u_{i*},>k}$ or not) and the reward is averaged over the hit ratio of all the pretend users in $\mathcal{U}_*^A$.

\textbf{Terminal.} The attacking process will stop when the number of actions reaches the budget $\triangle$. In addition, in the case when fewer user profiles are enough to successfully satisfies the promotion task, the process stops.



\subsection{ User Profile Selection via Hierarchical-structure Policy Gradient}


User profile selection aims to learn the strategy of selecting cross-domain user profiles. More specifically, it seeks to discover the set of users $\mathcal{U}^{B \rightarrow A} \subset \mathcal{U}^B$ that we can then inject their user profiles into the targeted recommender system's set of users $\mathcal{U}^A$ to achieve the goal of promoting a set of items. Here, the main challenges are how to handle a large-scale discrete action space (i.e., set of all user profiles) as well as achieve satisfied results under limited resources to interact with the target (black-box) recommender system $A$. Most existing RL techniques cannot handle such a large discrete action space problem as the time complexity of making a decision is linear to the size of action space~\cite{arulkumaran2017deep,zhao2018recommendations,zhao2019deep,dulac2015deep,chen2019large}. 
To address these challenges, we propose to utilize a hierarchical-structure policy gradient network with a masking mechanism to model the process of selecting a user profile (as shown in the left part in Figure~\ref{fig:framework}).  More specifically, we construct a  hierarchical clustering tree over cross-domain user profiles, where each leaf node is represented as a user profile and each non-leaf node is a policy network. Selecting a user profile in this hierarchical clustering tree is to seek a path from the root to a certain leaf of the tree.



\subsubsection{Hierarchical Clustering Tree over Cross-domain User Profiles}\label{sec:tree}
In the hierarchical clustering tree,  each leaf node is represented as a cross-domain user profile, while each non-leaf node is a policy network. However, the question remains how to construct the clustering tree. Hence, we propose to employ a top-down divisive approach that will repeatedly divide each cluster into small sub-clusters where leaf nodes under the same non-leaf node in the clustering tree should be more similar to each other than leaf nodes coming from another non-leaf node. We note that this process starts with the entire set of nodes at the root of the clustering tree. 

When constructing our hierarchical clustering tree we further add the constraint that it should be balanced to ensure the proper speedup (as an unbalanced clustering tree in the worst case could result in a linked list of policy networks on the order of the number of users). Hence, we use K-mean clustering method~\cite{lloyd1982least} and further modify it such that it forms clusters of equal size (off by at most a single user in size). To achieve this, at each non-leaf node when constructing the tree (top down), we first apply the traditional K-mean clustering on that current set of users to obtain the set of $c$ centroids. Note that the number of cluster centers (i.e., centroids) is set to as the same number of child nodes in the hierarchical clustering tree. Then, we reassign the users to these $c$ centroids one at a time based on their Euclidean distance to ensure we have a balanced set of clusters (in terms of their size). 

When constructing the clustering tree, one major consideration is how to balance the number of children per node against the height of the tree. To better understand this relationship between depth of hierarchical clustering tree $d$, the number of leaf node $|\mathcal{U}^B|$, and the number of child  node $c$,  we can observe the following:
\begin{align}
  c^{d-1} < |\mathcal{U}^B| = n^B \leqslant  c^d \nonumber
\end{align}
and the number of non-leaf nodes of the tree is $\mathcal{I} =  \frac{c^d - 1}{ c-1}$. In Section~\ref{sec:Experiments}, we perform an analysis of our proposed framework CopyAttack where we vary this balance between $c$ and $d$. 

We note that numerous features of the users' could be used for their representations, such as user attributes, review comments, and user-item interactions. In this work, we adopt the user-item interactions $\mathbf{Y}^B$ to represent the users because auxiliary information such as the user's attributes and review comments are not available. We use the user representations $\mathbf{p}^B \in \mathbb{R} ^{e}$ learned via matrix factorization (MF)~\cite{koren2009matrix} to measure similarity between users.

\subsubsection{Masking Mechanism}
While cross-domain user profiles contain informative signal of items, due to the limited number of queries in the target recommender system, not all the cross-domain user profiles are useful for attacking a specific target item. Actually, only user profiles related to the specific target items would be useful. Therefore, we need to tune the hierarchical clustering tree  with a masking mechanism to
locate some percentage of related cross-domain user profiles for the target items. More specifically, for each target item,  we take an approach of masking the cross-domain user profiles that do not include the target item.  As shown in the left part in Figure~\ref{fig:framework},  the path from non-leaf node $3$ to node $7$ is masked, since the cross-domain profiles of user $u_7^B$ and $u_8^B$ do not include the target item (with pink color). As such, these cross-domain user profiles (i.e., $u_7^B$ and $u_8^B$) can not be explored by the RL agent, which might further help reduce the action space. This reduction in the action space in turn is efficient then to locate useful cross-domain user profiles to perform an effective attack. We again note that the target item $v_*$ comes from the set $\mathcal{V} = \mathcal{V}^A \cap \mathcal{V}^B$. In other words, the target item exists in the source domain so the masking will never result in the entire tree being masked. 

\subsubsection{Hierarchical-structure Policy Gradient}

With the hierarchical clustering tree, the purpose of user profile selection is to learn the policy $p(a_t^u|s^u_t)$ for seeking a path $a^u_t$ from the root to a certain leaf of tree (i.e., user in $\mathcal{U}^B$) at state $t$.   Each non-leaf node in the tree is a policy gradient network, which can be modeled as a Multi-Layer Perceptron (MLP). As such, there are $\mathcal{I}$  policy gradient networks with $\theta=\left \{ \theta_1, \theta_2,..., \theta_\mathcal{I}    \right \}$ in  the hierarchical clustering tree. 

In particular, the policy network at $node_i$ (having MLP parameters denoted as $\theta_i$) first takes the current state as input and outputs a probability distribution over all child nodes of $node_i$. Then, one of the children is selected to move based on the probabilities. The selection process then keeps moving down the clustering tree of policy networks until reaching a leaf node (i.e., a user profile), which can form the path of length $d$ from the root to the leaf node as follows:
\begin{align}
   a^u_t=\left \{ a_{[t,1]}^u, a_{[t,2]}^u,..., a_{[t,d]}^u    \right \} \nonumber
\end{align}
 
This selection process can be decomposed to multiple steps according to selected path $a^u_t$ as follows:
\begin{align}
    p^u(a_t^u|s^u_t) &=\prod^d p^u_d (a_t^u|\cdot, s^u_t) \nonumber \\
 \nonumber   &=  p^u_{d} (a_{[t,d]}^u| s^u_t)  \cdot  p^u_{d-1} (a_{[t,d-1]}^u| s^u_t)  \cdots  p^u_{1} (a_{[t,1]}^u|s^u_t) \nonumber
\end{align}

We represent the state $s^u_t$ with the target item $v_*$ and previous selected users $\mathcal{U}_t^{B \rightarrow A} = \left \{  u^B_1, ..., u^B_i, ..., u^B_{t} \right \}$. We combine them together with a Multi-Layer Perceptron  (MLP).   To decide which path we will move to, by estimating the probability distribution over the children at $node_i$ (i.e., the policy network parameterized by $\theta_i$), as follows:
\begin{align}
    \mathbf{x}_{v_*} &= RNN(\mathcal{U}_t^{B \rightarrow A}), \nonumber \\
    p^u_i(\cdot | s^u_t) &= softmax (MLP( [\mathbf{q}^B_{v_*}  \oplus  \mathbf{x}_{v_*}] | \theta_i^u)) \nonumber
\end{align}
where $\mathbf{q}^B_{v_*}  \in \mathbb{R} ^{e}$ is the pre-train item representation via Matrix Factorization (MF) coming from the source domain $B$. We model the selected users $\mathcal{U}_t^{B \rightarrow A}$ at state $s_t$ with an RNN model and denote its representation as $\mathbf{x}_{v_*}$. Here we use $\oplus $ to denote the concatenation operation. Also, here we seed the process by selecting action $a^u_0$ (i.e., the first user to inject in the target recommender system) at random, since at that time $\mathcal{U}_0^{B \rightarrow A}$ is empty and would not provide any insights from the RNN. We leave it as future work to investigate other methods of seeding this process, although a random action is one commonly performed in practice.

An illustration example of the process of selecting cross-domain user profiles is shown in the left part in Figure~\ref{fig:framework}. We have 8 user profiles, and build a balanced hierarchical clustering tree with depth 3 over user profiles in the source domain $B$. For a given state $s_t$, the status point is initially located at the root ($node_1$), and moves to one of its child nodes to ($node_2$) according to the probability distribution given by the policy network \emph{PN-1} corresponding to the root ($node_1$). The process of selecting can stop when the state point arrives at a leaf node in the tree; in this case, user $u_3^B$'s profile. Note that at the state point $node_5$, the path from $node_5$ to leaf node $u_4^B$ is masked since the profile of source domain user $u_4^B$ does not include the currently attacking target item. The example path for this selection is $a^u_t=\left \{ node_1, node_2, node_5, u_3^B \right \}$, as the path with  green color in the figure.

Although we now have an efficient mechanism for selecting the set of source domain users that the attacker will copy into the target domain, we again note here that there could be some problems with directly copying these nodes. It could be the case that not all items in a user's profile are useful in the promotion attack and could just inject noise and/or increase the attack cost. Hence, next we will introduce another policy gradient network that will learn how to craft user profiles by reducing the number of items in the user's profile (i.e., the items they have interacted with).








\subsection{User Profile Crafting}





Not all the interactions towards items in cross-domain user profiles are helpful.
Directly injecting the raw user profiles into the target recommender system
may lead to increase the attacking budgets and include some noise.  To address this challenge, we propose a clipping operation to craft the raw user profiles via policy network, as shown in the middle of Figure~\ref{fig:framework}.

More specifically, we first discrete the length into 10 different levels as follows,
\begin{align}
W=\left \{ 10\%, 20\% , 30\% , 40\% , 50\% , 60\% , 70\% , 80\% , 90\% , 100\%   \right \} \nonumber
\end{align}
Then, a policy network is introduced to choose the action $a^l=w $ from the set $W$ to decide the length we keep (i.e., number of interactions for that selected user profile). As the raw selected user profile includes the target item $v_*$,  the raw user profile is clipped around the target item with the window size $w$. As such, we can consider the forward and backward related items.  For example, the selected raw profile of user $u_i^B$ with 10 items is as follows, 
\begin{align}
    P_{u_i}^B= \left \{ v_1\rightarrow   v_2\rightarrow   v_3\rightarrow   v_4\rightarrow    v_{5*}\rightarrow   v_6\rightarrow   v_7\rightarrow   v_8\rightarrow   v_9\rightarrow   v_{10}  \right \} \nonumber
\end{align}
If the policy network takes the action $a^l=50\%$, the new user profile through  the clipping operation can keep around $50\%$ raw user profile as follows:
\begin{align}
    \hat{P}_{u_i}^B= \left \{ v_3\rightarrow v_4\rightarrow    v_{5*}\rightarrow v_6\rightarrow v_7 \right \} \nonumber
\end{align}

The state $s^l_t$ for model clipping operation can be decided by the selected user $u_i$ and target item $v_*$. We estimate the probability of choosing action $a^l$ over the set $W$ with the state $s_t^l$, as follows,
\begin{align}
    p^l(\cdot | s^l_t) &= softmax (MLP( [\mathbf{p}^B_{i}  \oplus  \mathbf{q}^B_{v_*}] |\theta ^l)) \nonumber
\end{align}
where $ \mathbf{p}_i^B \in \mathbb{R} ^{e}$ and  $ \mathbf{q}_{v*}^B \in \mathbb{R} ^{e}$ are the pre-trained user and item representations via MF in source domain, respectively. Also, we note that when considering how to craft the user profiles there are perhaps a few options that could be taken on how to utilize $a^l$ for reducing the user profile size. For example, intuitively randomly selecting a subset to keep would not make sense due to the fact it would lose the temporal relations of items that were interacted by the given user around the same time as the target item. Furthermore, if we were to select perhaps based on the most similar nodes to the target node from the user's profile, then this might result in a less realistic user profile that could potentially more easily be detected. Hence, our selection of clipping the user profile with a window size $w$ around the target item indeed appears to be the logical mechanism for clipping.




\subsection{Injection Attack and Queries}

To perform attacking in the black-box setting, we only have query access to the target model and can get query feedback consisting of Top-$k$ recommended items for specific users. Hence, in CopyAttack's last stage we actually inject the selected user profiles that we have crafted from the source domain to the target domain. Then, once injected, the attacker can utilize their set of pretend users $\mathcal{U}_*^A$ they have already established in the target domain to gauge the effectiveness of the injected user profiles and define a corresponding reward. More specifically, here we use the reward function defined in Eq.~(\ref{eq:reward}) where the effectiveness is defined based on the hit ratio (HR@K) of the targeted item $v_*$ aggregated over the set of pretend users' (i.e., those in the set $\mathcal{U}_*^A$) Top-$k$ recommendations. We note that these Top-$k$ recommendations are the result/feedback upon performing queries of target system $A$. Once obtaining the reward it is then used to update the policy networks for both the profile selection and profile crafting CopyAttack components. 

%% file: experiments.tex
\section{Experiment}
\label{sec:Experiments}
In this section, we conduct experiments to verify the effectiveness of our model. 
We first introduce the experimental settings, then discuss the results (i.e.,  performance comparison) of various baselines, and finally study the impact of different components in our model.

\subsection{Experimental Settings}
\subsubsection{Datasets}

We have used two cross-domain real-world datasets in our experiments to validate the performance of CopyAttack.

\begin{table}[]

\small 
	\setlength\tabcolsep{5.0pt} 
    \renewcommand{\arraystretch}{0.9} 
\caption{Statistics of Two Datasets}
\label{tab:dataset}
\begin{tabular}{c|c|c|c}
\hline
\multicolumn{2}{c|}{Datasets (Target, Source)}                                                                                                                 & (ML10M, Flixster) & (ML20M, Netflix) \\ \hline
\multirow{3}{*}{\begin{tabular}[c]{@{}c@{}}Target\\  Domain\end{tabular}} & \# of Users                                                        & 19,267         & 38,087        \\ \cline{2-4} 
                                                                          & \# of Items                                                        & 6,984          & 8,325         \\ \cline{2-4} 
                                                                          & \# of Interactions                                                  & 437,746        & 838,491      \\ \hline
\multirow{3}{*}{\begin{tabular}[c]{@{}c@{}}Source \\ Domain\end{tabular}} & \# of Users                                                        & 93,702         & 478,471      \\ \cline{2-4} 
                                                                          & \begin{tabular}[c]{@{}c@{}}\# of Overlapping\\  Items\end{tabular} & 5,815          & 5,193         \\ \cline{2-4} 
                                                                          & \# of Interactions                                                 & 4,680,700      & 62,937,958   \\ \hline
\end{tabular}
\end{table}

\textbf{MovieLen10M\footnote{https://grouplens.org/datasets/movielens/10m/} $\&$ Flixster\footnote{https://sites.google.com/view/mohsenjamali/home} (ML10M-FX)}. Both datasets are popular online platforms for movie recommendation services, in which they have millions of movies. Users in these two platforms can watch them and give their personal comments (e.g., rating). 
Here, we take Movielen10M (ML10M) dataset as the target domain, which is utilized to be attacked. Flixster (FX) dataset is treated as the source domain to be used to copy some user profiles to attack the Movielen10M (ML10M) domain. In these two datasets, they have a lot of items in common, where overlapping items can be aligned by the movie names. 
We only keep the interactions that have a rating score of $5$. After filtering, this cross-domain dataset (ML10M-FX) has 5,815 overlapping items.
 
\textbf{MovieLen20M\footnote{ https://grouplens.org/datasets/movielens/20m/}  $\&$ Netflix\footnote{https://www.kaggle.com/laowingkin/netflix-movie-recommendation}  (ML20M-NF)}. These two datasets are also online platforms for movie recommendation services. We take Movielen20M (ML20M) dataset as the target domain and Netflix (NF) is the source domain. 
We identify movies with the same name and the published year. We then perform filtering operations similar to the ML10M-FX dataset. In this cross-domain dataset we have 5,193 overlapping items. 

The statistics of these datasets are presented in Table~\ref{tab:dataset}. Note that we only keep the overlapping items in the source domain. 

\subsubsection{Evaluation Metrics}

In order to evaluate the quality of the recommender systems, we use two popular accuracy metrics for Top-K recommendation~\cite{he2017neural}: Hit Rate (HR@K) and Normalized Discounted Cumulative Gain (NDCG@K). We set $K$ as 20, 10, and 5. Higher values of the HR@K and NDCG@K indicate a better predictive performance. As the ranking task is too time-consuming to rank all the items for all the users, we  randomly sample 100 items that the user did not interact with and then rank the test item among them.

\begin{table*}[]
\small
	\setlength\tabcolsep{5.0pt} 
    \renewcommand{\arraystretch}{1.0} 
\caption{ Performance comparison of different attacking methods for recommender systems}
\label{tab:performance}
\begin{tabular}{|c|c|c|c|c|c|c|c|c|}
\hline
\textbf{Dataset}                          & \textbf{Algorithms} & \textbf{HR@20}  & \textbf{HR@10}  & \textbf{HR@5}   & \textbf{NDCG@20} & \textbf{NDCG@10} & \textbf{NDCG@5} & \textbf{\begin{tabular}[c]{@{}c@{}}\# Average Items\\  per User Profile\end{tabular}} \\ \hline
\multirow{9}{*}{\textbf{ML10M- FX}} & Without Attack      & 0.0378          & 0.0228          & 0.0220           & 0.0231           & 0.0195           & 0.0192          & 0                                                                                     \\ \cline{2-9} 
                                          & RandomAttack       & 0.0391          & 0.0230           & 0.0222          & 0.0233           & 0.0195           & 0.0192          & 46                                                                                    \\ \cline{2-9} 
                                          & TargetAttack40     & 0.1203          & 0.0583          & 0.0094          & 0.0353           & 0.0195           & 0.0041          & 495                                                                                    \\ \cline{2-9} 
                                          & TargetAttack70    & 0.1772          & 0.0854          & 0.0354          & 0.0569           & 0.0341           & 0.0181          & 818                                                                                   \\ \cline{2-9} 
                                          & TargetAttack100   & 0.1166          & 0.0520           & 0.0226          & 0.0369           & 0.0209           & 0.0114          & 1350                                                                                  \\ \cline{2-9} 
                                          & PolicyNetwork      & 0.1936          & 0.0665          & 0.0250           & 0.0570            & 0.0258           & 0.0126          & 705                                                                                   \\ \cline{2-9} 
                                          & CopyAttack-Masking     & 0.0376         & 0.0227         & 0.0220         & 0.0230          & 0.0195           & 0.0192         & 49                                                                                    \\ \cline{2-9} 
                                          & CopyAttack-Length   & 0.0857         & 0.0434         & 0.0198         & 0.0282          & 0.0177          & 0.0101         & 1280                                                                                  \\ \cline{2-9} 
                                          & \textbf{CopyAttack} & \textbf{0.2596} & \textbf{0.1103} & \textbf{0.0415} & \textbf{0.0799}  & \textbf{0.0425}  & \textbf{0.0205} & 695                                                                         \\ \hline
\multirow{9}{*}{\textbf{ML20M-NF}}   & Without Attack      & 0.0461          & 0.0043          & 0.0000               & 0.0115           & 0.0013           & 0.0000               & 0                                                                                     \\ \cline{2-9} 
                                          & RandomAttack       & 0.0468          & 0.0050           & 0.0000           & 0.0118           & 0.0015           & 0.0000               & 124                                                                                   \\ \cline{2-9} 
                                          & TargetAttack 40    & 0.1016          & 0.0405          & 0.0056          & 0.0288           & 0.0133           & 0.0024          & 203                                                                                   \\ \cline{2-9} 
                                          & TargetAttack70    & 0.1006          & 0.0402          & 0.0054          & 0.0285           & 0.0132           & 0.0023          & 321                                                                                   \\ \cline{2-9} 
                                          & TargetAttack100   & 0.0581          & 0.0006          & 0.0000               & 0.0139           & 0.0002           & 0.000               & 593                                                                                   \\ \cline{2-9} 
                                          & PolicyNetwork      & --            & --            & --            & --             & --             & --            & --                                                                                  \\ \cline{2-9} 
                                          & CopyAttack-Masking     & 0.0500          & 0.0045        & 0.0000        & 0.0125        & 0.0001        & 0.0000               & 133                                                                                   \\ \cline{2-9} 
                                          & CopyAttack-Length   & 0.0655         & 0.0018         & 0.0000              & 0.0158          & 0.0005          & 0.0000               & 496                                                                                   \\ \cline{2-9} 
                                          & \textbf{CopyAttack} & \textbf{0.2704} & \textbf{0.124}  & \textbf{0.0797} & \textbf{0.0969}  & \textbf{0.0609}  & \textbf{0.0467} & 255                                                                          \\ \hline
\end{tabular}
\end{table*}

\subsubsection{Attacking Environment and Parameter Settings}
Graph Neural Networks (GNNs) based techniques are the state-of-the-art models for recommender systems~\cite{wang2019neural}. 
The popular GNNs model in recommendations, PinSage, for item recommendations~\cite{ying2018graph}, which aggregates the local neighbors (users/items) in an inductive way, has been applied in industry~\cite{ying2018graph,hamilton2017inductive}. Therefore, we adopt this model as our target model, where user and items representations can be learned via aggregating their local neighbors (items/users).   

To train this target recommender systems, we randomly split the target domain datasets, where we have $80\%$ as a training set for learning the parameters, $10\%$ as a validation set to tune hyper-parameters, and $10\%$ as the test set. For all neural network methods, we randomly initialized model parameters with a Gaussian distribution, where the mean and standard deviation is 0 and 0.1, respectively. The learning rate and embedding size are set to be 0.001 and 8. The early stopping strategy was performed, where we stopped training if the HR@10 on validation set increased for 5 successive epochs.  After completing training, the final performance on testing datasets is 0.549 with HR@10 metrics for ML-10M dataset, and 0.5474 for ML-20M dataset. After training, the recommender systems in the target domain are fixed, where the model structure and model parameter can not be changed.
Then, we use the well-trained model in a black-box attacking environment and evaluate the different attacking performance. 

We randomly sample 50 target items with less than 10 interactions to perform attacking on the target domain. Without being specifically mentioned, the main budget for attacking is the number of cross-domain profiles, where we set the maximum budget as 30. The number of pretend user in  $\mathcal{U}^A_*$ is set to 50 for both datasets. To get the feedback from the target system, we perform queries on the target system after each 3 injections.


We implemented the proposed method on the basis of Tensorflow. 
The learning rate, the size of action, and discount factor are set to 0.001, 8, and 0.6, respectively. The hierarchical clustering tree is set to 3 layers for Flixster dataset and 6 layers for Netflix dataset. The user and item representation is trained with Matrix Factorization techniques, where we use same hyper-parameters to train (learning rate, embedding, etc).


\subsubsection{Baselines}
Most of existing attacking methods in recommender systems are under white-box setting, where they assume the attack can have full knowledge of the target model (e.g., model structure, parameters) and access the datasets. There is not existing black-box attack for recommender systems. We build some baselines to evaluate the performance of attacking as follows:

\textbf{ RandomAttack}: This baseline is proposed to  randomly sample cross-domain user profiles to attacking the target recommender systems.
\textbf{ TargetAttack40}:  Rather than randomly sampling user profiles from source domain, this baseline is to sample the user profile from the source domain with the target item which is going to be attacked. Moreover, we apply the user profile crafting operations as our proposed model to reserve 40$\%$ of user profiles.
\textbf{TargetAttack70}: This baseline is similar with TargetAttack40, while setting the length of user profile as 70$\%$.
\textbf{TargetAttack100}: This  method is used to directly random sample user profiles including target items from source domain,  without further crafting the selected user profile as TargetAttack40 and TargetAttack70.

In addition, we also build some baselines based on our proposed methods as follows:

\textbf{PolicyNetwork}: This method directly uses the policy gradient on the action space, without considering the hierarchical clustering tree.  
\textbf{CopyAttack-Masking}: This method is used to evaluate the effectiveness of masking mechanism in our proposed framework. In other words, the attack can select any user profile in the source domain. Note that the user profile crafting operation in  this baseline is also be removed, since the attack has larger probability to select the user profile without the target items. 
\textbf{CopyAttack-Length}: This method is used to evaluate the effectiveness of user profile crafting operation in our proposed framework, where we remove the user profile crafting operation.   
\begin{figure*}[t]
\centering
{\subfigure[10M HR@20]
{\includegraphics[width=0.2345\linewidth]{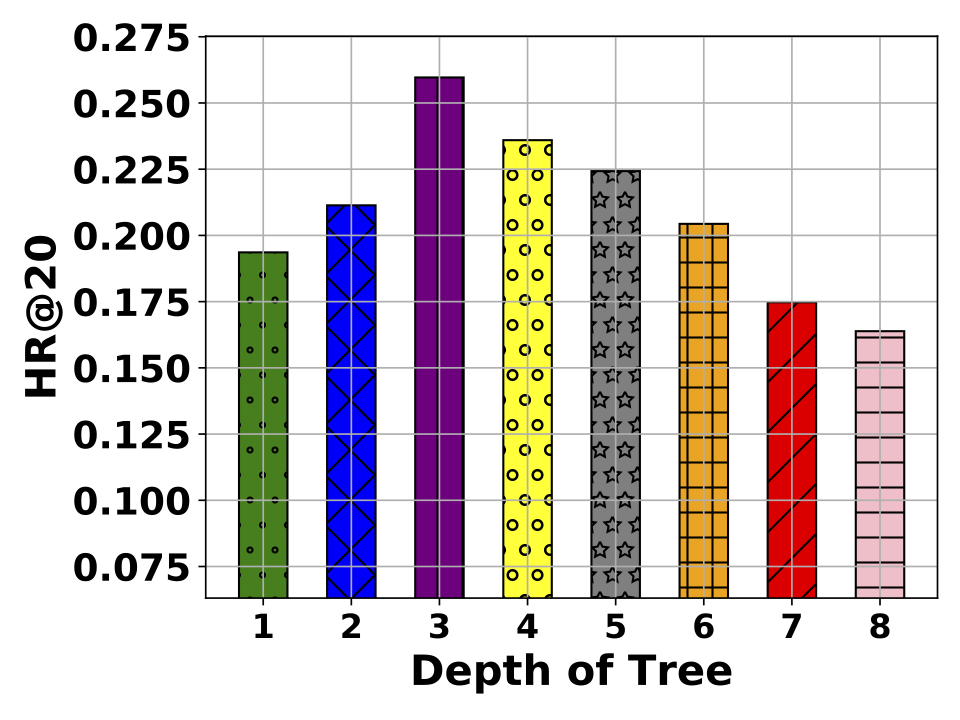}}}
{\subfigure[10M NDCG@20]
{\includegraphics[width=0.2345\linewidth]{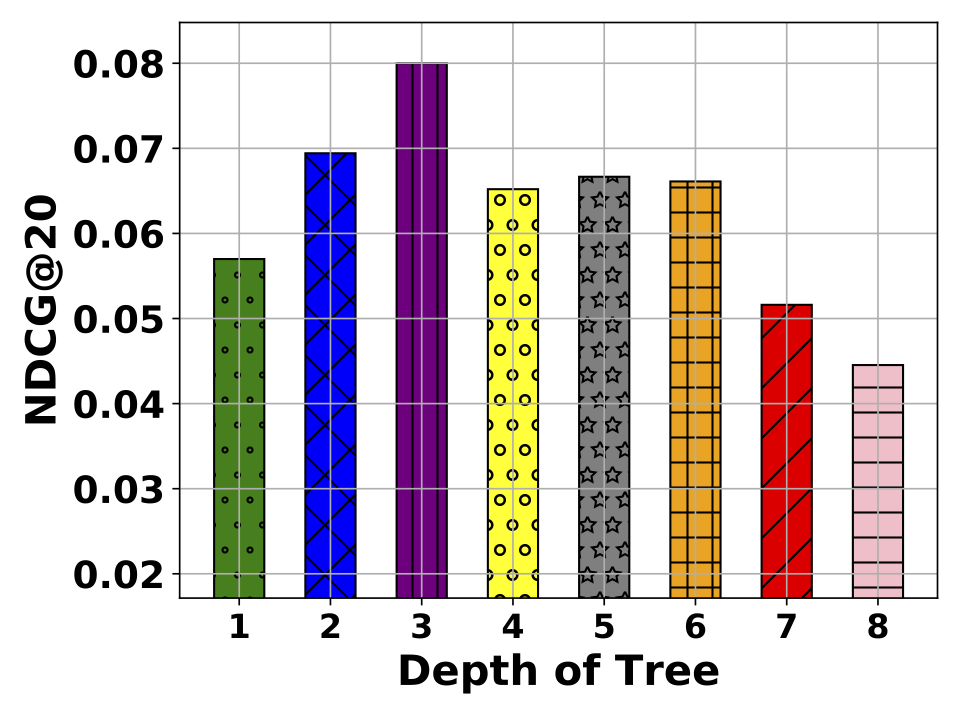}}}
{\subfigure[20M HR@20]
{\includegraphics[width=0.2345\linewidth]{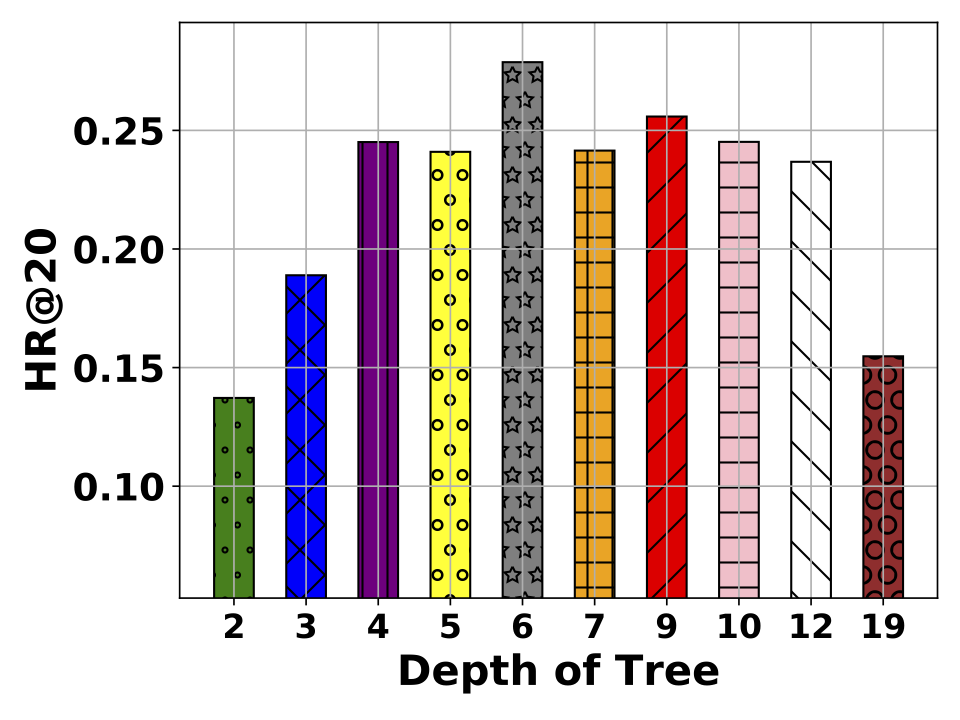}}}
{\subfigure[20M NDCG@20]
{\includegraphics[width=0.2345\linewidth]{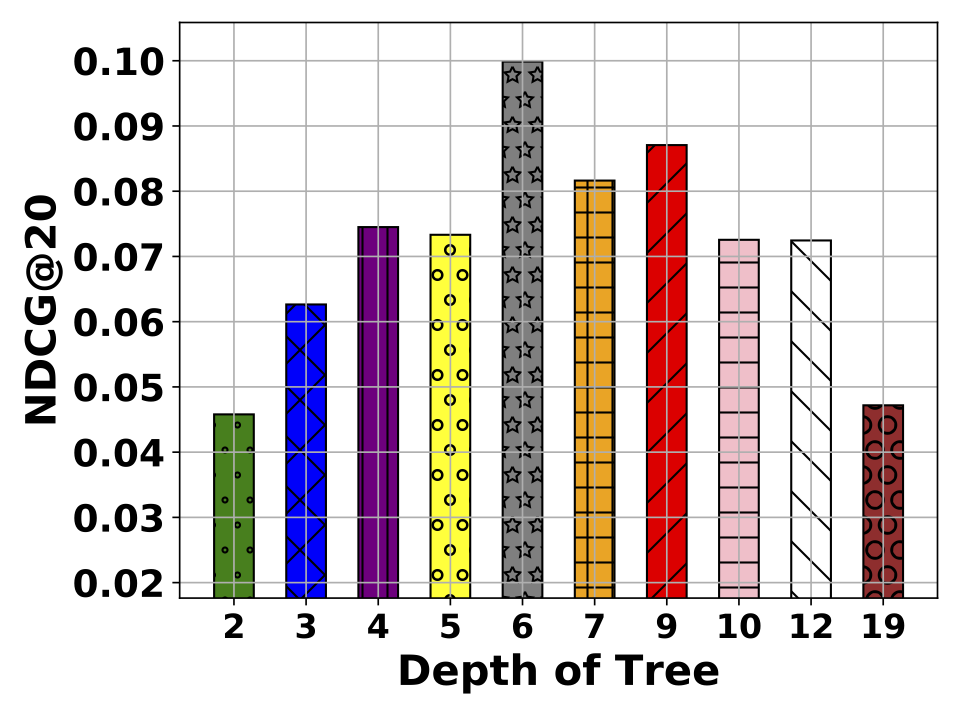}}}
\caption{Effect of Depth on Hierarchical Clustering Tree. }\label{fig:Depth}
\end{figure*}

\begin{figure*}[t]
\centering
{\subfigure[10M HR@20]
{\includegraphics[width=0.2345\linewidth]{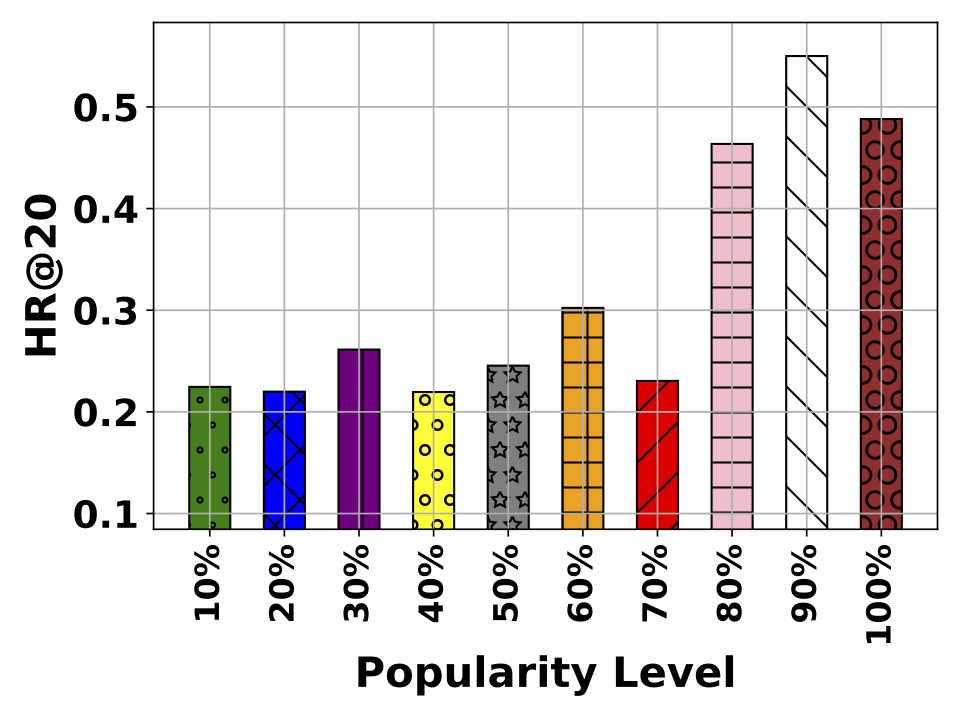}}}
{\subfigure[10M NDCG@20]
{\includegraphics[width=0.2345\linewidth]{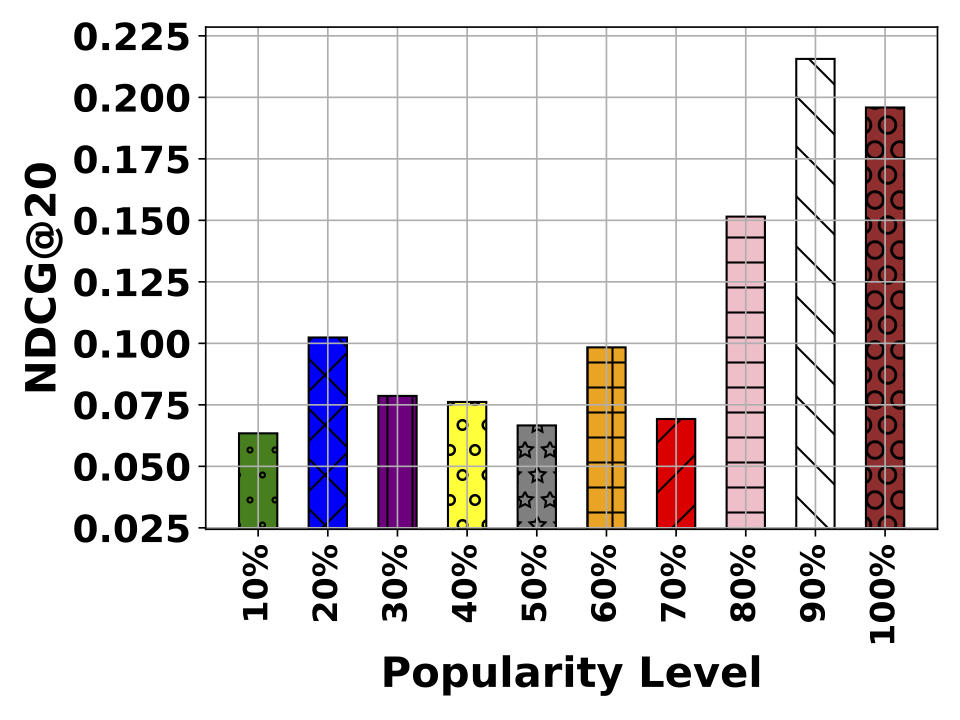}}}
{\subfigure[20M HR@20]
{\includegraphics[width=0.2345\linewidth]{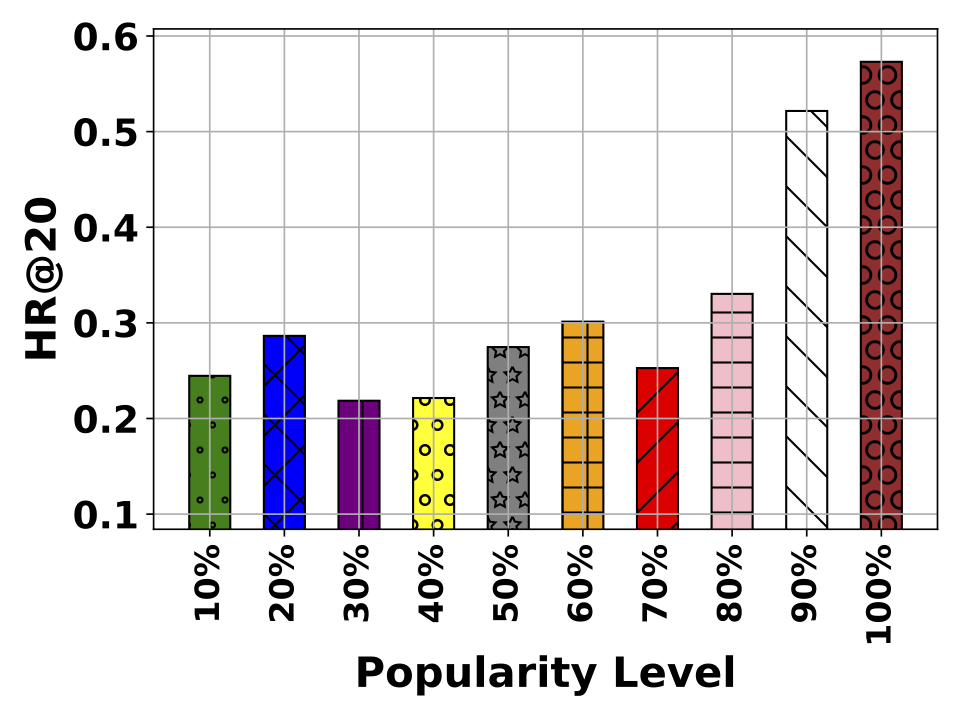}}}
{\subfigure[20M NDCG@20]
{\includegraphics[width=0.2345\linewidth]{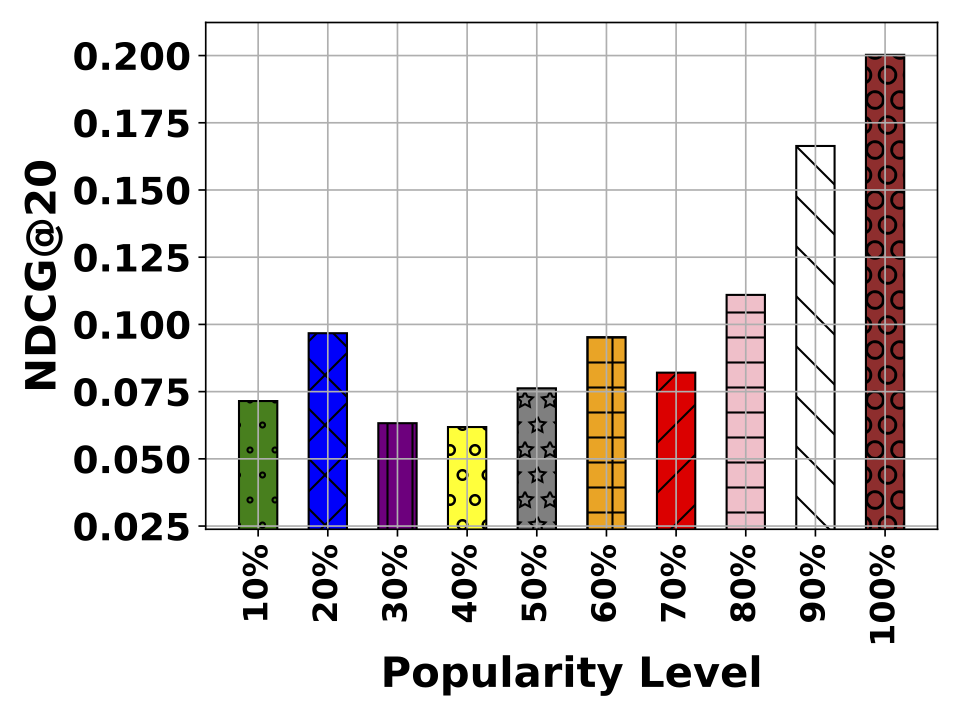}}}
\caption{Effect of Item Popularity.}\label{fig:popularity}
\end{figure*}


\begin{figure*}[t]
\centering
{\subfigure[10M HR@20]
{\includegraphics[width=0.4585\linewidth]{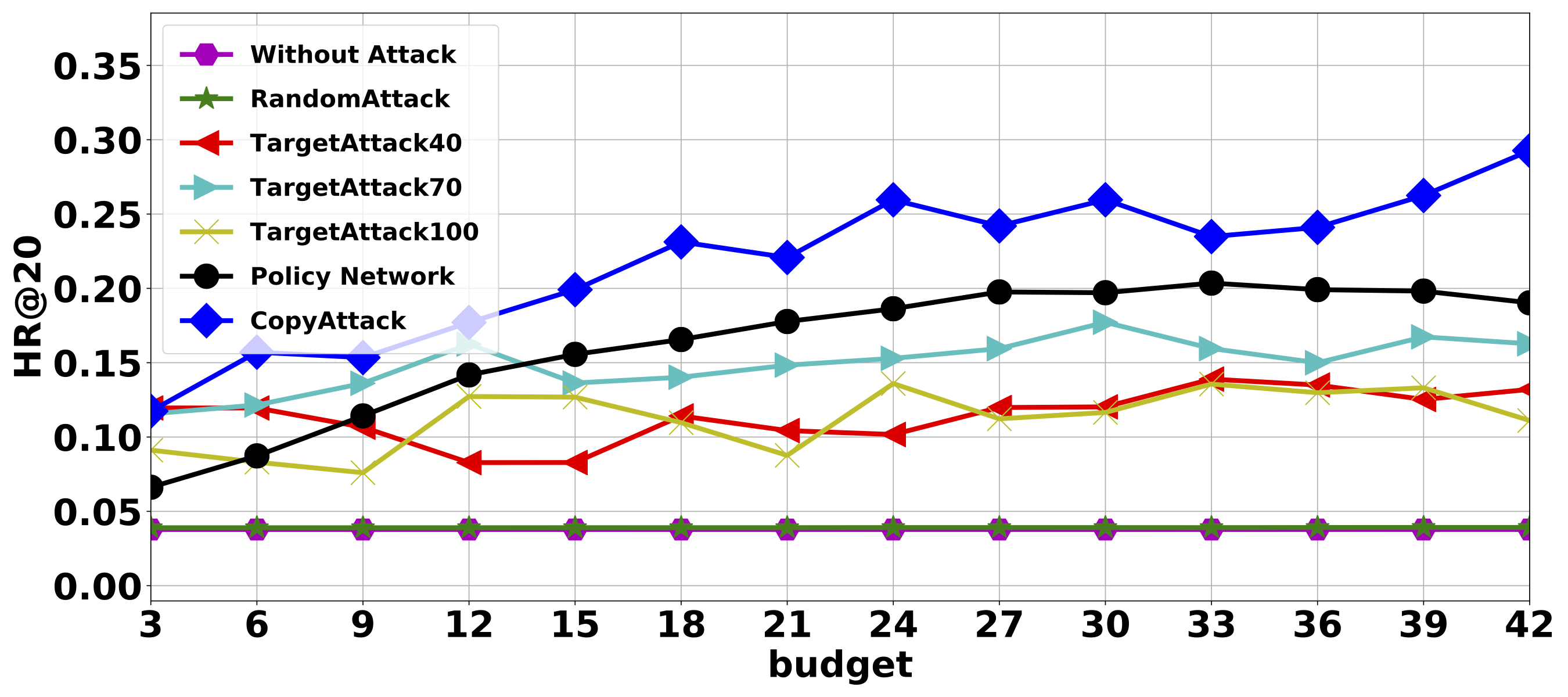}}}
{\subfigure[10M NDCG@20]
{\includegraphics[width=0.4585\linewidth]{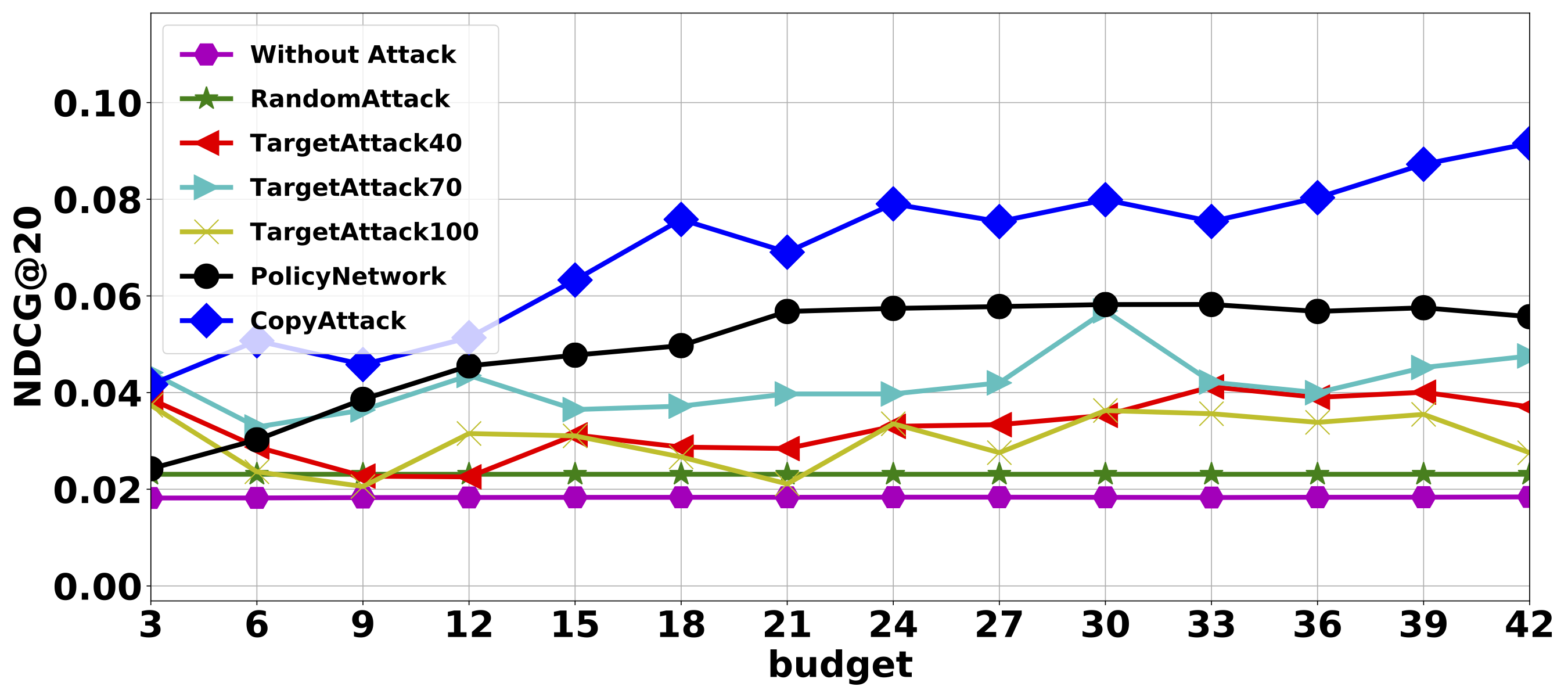}}}
\caption{Effect of Budget (Cross-domain User Profiles)  on ML10M-FX.}\label{fig:budget10M}
\end{figure*}

\subsection{Performance Comparison of Recommender Systems}
We first compare the attacking performance of all methods.  Table~\ref{tab:performance} shows the overall attacking performances on different methods $w.r.t$ HR@K and NDCG@K on ML10M-FX and ML20M-NF datasets. We have the following main findings.

Randomly sampling cross-domain user profiles without any strategies can not help promote the target items. When sampling user profiles with the sampling strategy where the user profiles should include the target items, the performance can be improved significantly. In addition, when we constrain the sampling cross-domain user profile scope into the users who include the target items, this kind of method can obtain much better performance. This indicates the user profiles with the target item are informative to help perform attacking.

When considering the length of cross-domain user profiles, the methods without target item constraint have very low item budget (less than 50).  When harnessingg this constraint on different TargetAttack-(40, 70, 100), we found that the methods on user profile without crafting perform the worse. It implies that introducing the user profile crafting is important. We will further analyse the budget from the number of user profile perspective in next section.

To better understand CopyAttack, we compare with PolicyNetwork, CopyAttack-Masking, and CopyAttack-Length. We can see that, for PolicyNetwork method, the performance of CopyAttack degrades when eliminating the effect of the hierarchical clustering tree. Note that PolicyNetwork method  on ML20M-NF does not work, since we can not obtain its results in 48 hours, while we can obtain the results of others in just few hours. These observations suggest the power of  the hierarchical clustering tree. We also further study the impact of the hierarchical clustering tree on next section.
Meanwhile, when we remove the user profile crafting component,  the promotion performance decrease too much and the item budget is very huge, since the selected user profiles might introduce too much noise and degrade the performance. Moreover, when the masking mechanism is removed upon the CopyAttack-Length, CopyAttack-Masking performs much worse. These results support that the masking mechanism and user profile crafting component are beneficial to select strong user profiles and reduce the item budget for each user profile.  

\subsection{Model Analysis}

In this subsection, we study the impact of model components and model hyper-parameters.

\subsubsection{Effect of Depth on Hierarchical Clustering Tree. }
The hierarchical clustering tree, as discussed in Section~\ref{sec:tree}, is investigated here where we have shown the performance when varying the depth of the tree (i.e., the value of $d$). We can observe in Figure~\ref{fig:Depth} that for 20M $d=3$ performs the best in terms of HR@20 and NDCG@20. Similarly, in 20M $d=6$ performs the best. The reason for this is believed to be the trade of in terms of how detailed the clusters can be and the number of policy networks. This is because the deeper the tree we have more policy networks that need to be learned. In comparison, shallower trees have less policy networks, but can harness the efficiency in terms of run-time and ability to have a few large clusters to guide the source user profile selection.

\subsubsection{Effect of Item Popularity. }
In this section, we study what kinds of items are vulnerable to attack.  To achieve it, we group the item in target domain based on their popularity. Specifically, we have 10 different groups, where each group account for $10\%$ of items in target domain. We then sample 50 target items from these 10 different groups respectively. At last, we evaluate the performance on them. Th results are given on Figure~\ref{fig:popularity}. We note that the target items with high popularity can be vulnerable to attack, where the top $30\%$ of items are vulnerable. 

\subsubsection{Effect of Budget (Cross-domain User Profiles). }

To perform attacking under black-box attack,  the budget is very important. In this section, we investigate how the budget affect the performance on different attacking methods. Figure~\ref{fig:budget10M}  show the performance with varied budget on ML10M-FX  dataset. We first note, the RandomAttack remains stable not matter how many user profiles. When the value of budget increase, the performance of methods injecting user profile with target items tends to increase first. And then TargetAttack40, TargetAttak70, and TargetAttack100 can not keep increasing when budget becomes too large, while CopyAttack keep  increasing  since this method perform queries and get more and more reward to train the attack. The results on ML20-NF is shown at Supplementation Section.  


%% file: conclusion.tex
\vspace{-1.5ex}
\section{Conclusion and Future work}
\label{sec:conclusion}

Many user-oriented online services make use of deep learning based recommender systems to suggest personalized lists for users to interact with. Although works have shown that these models are susceptible to attack, more recent studies have shown that state-of-the-art defense strategies are able to detect data poisoning attacks in recommender systems. This is primarily due to the fact that injected fake user profiles are easily detected. Hence, in this work we have proposed a cross-domain approach to copy users from a source domain to the target domain towards the goal of promoting certain target items. More specifically, we have introduced a reinforcement learning based black-box approach that makes use of policy gradient networks to first select users to copy, refines/crafts their profiles, and finally injects them in the target domain where we can then observe some feedback in terms of Top-$k$ recommendations on our set of pretend users. These pretend users are then used to determine the reward for updating our model parameters.

Our thorough experiments on two real-world datasets show the superiority of the proposed framework, CopyAttack, over a set of competitive baselines. Then, we furthermore performed model analysis to better understand the behavior of CopyAttack. Our future work will be towards effective strategies for targeted attacks on items that need not be in the source domain and also for demotion and furthermore include more rich side information.

%% file: appendix.tex
\section{Supplementation}

\begin{figure}[!h]
\centering
{\subfigure[20M HR@20]
{\includegraphics[width=0.95\linewidth]{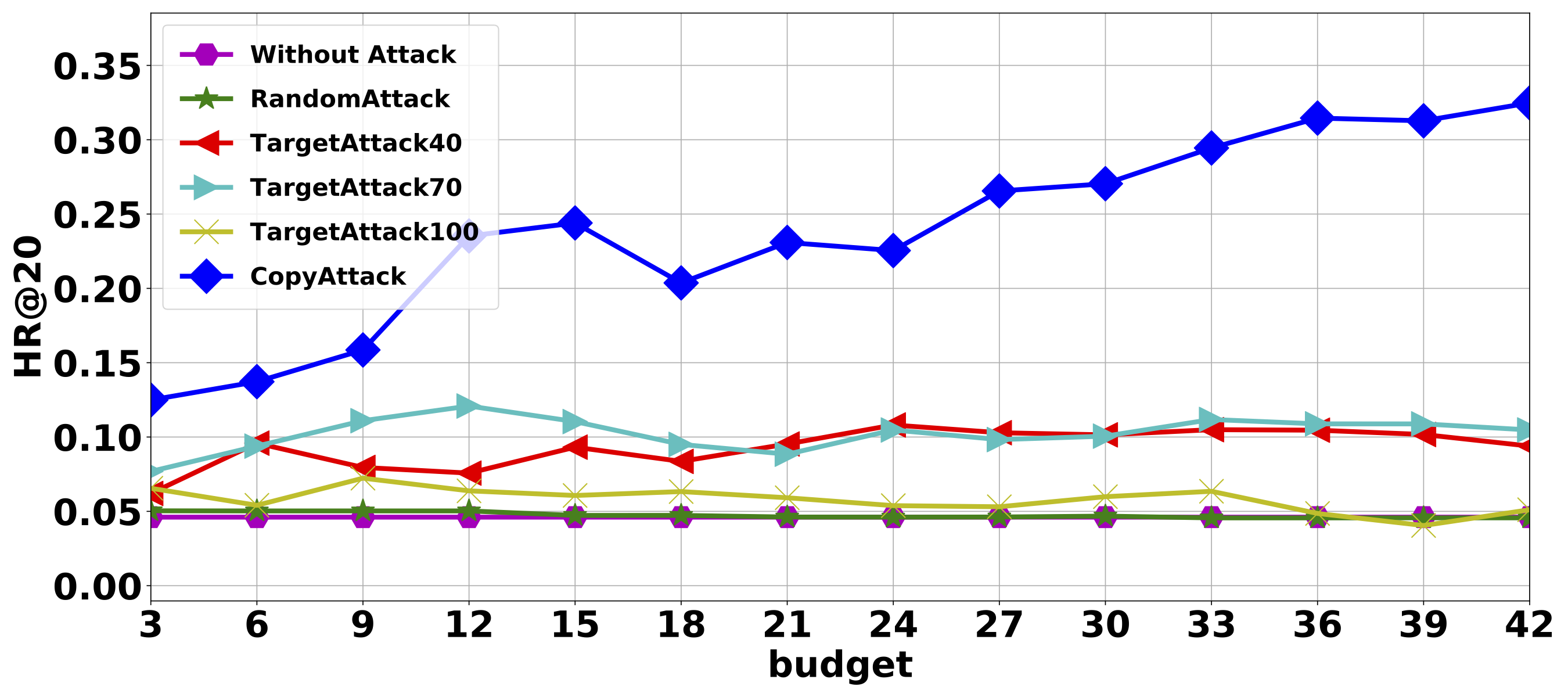}}}
{\subfigure[20M NDCG@20]
{\includegraphics[width=0.95\linewidth]{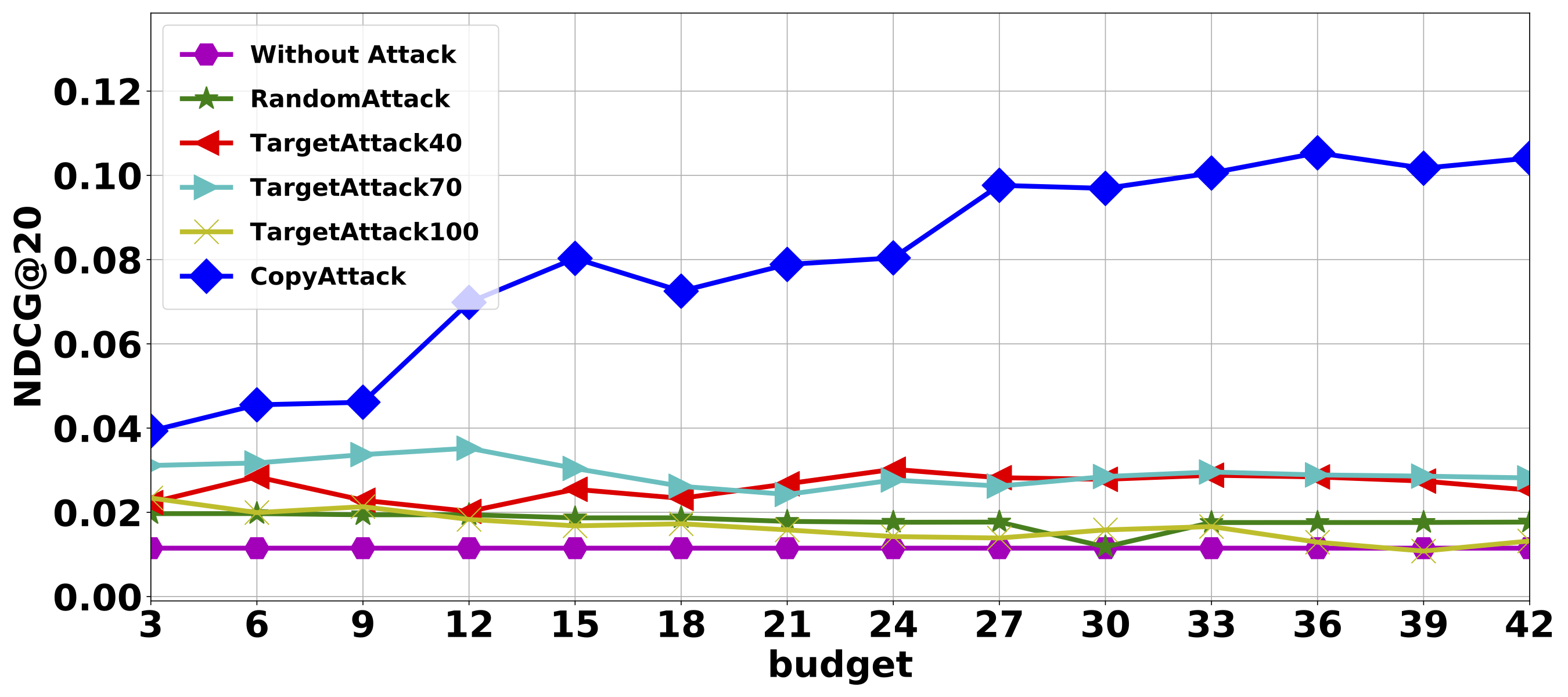}}}
\caption{Effect of Budget (Cross-domain User Profiles) on ML20M-NF.}\label{fig:extra}
\end{figure}

We include the experiment result about the effect of budget for understanding our proposed method in Figure~\ref{fig:extra}. We again note that in this dataset the PolicyNetwork baseline was unable to finish in a reasonable time limit of 48 hours, so we do not report their performance. This also further strengthens the usefulness of the hierarchical clustering tree as compared to a single policy gradient network for the entire action space of all users (in the source domain), since CopyAttack obtains the results in just a few hours (e.g., ~3 hours). Please note that we will release our code upon the acceptance of this paper for reproducibility.  

